\documentclass[reprint,amssymb, amsmath, aps, superscriptaddress, showpacs, footinbib,  prb]{revtex4-1}
\usepackage{graphicx}
\usepackage{color}

\newcommand{\eff}{\text{eff}}
\newcommand{\AFM}{\text{AFM}}
\newcommand{\FM}{\text{FM}}

\begin{document}

\title{Superposition of ferromagnetic and antiferromagnetic spin chains\\ in the quantum magnet BaAg$_2$Cu[VO$_4]_2$}

\author{Alexander A. Tsirlin}
\email{altsirlin@gmail.com}
\affiliation{Max Planck Institute for Chemical Physics of Solids, N\"{o}thnitzer Str. 40, 01187 Dresden, Germany}

\author{Angela M\"oller}
\email{amoeller@uh.edu}
\affiliation{Texas Center for Superconductivity, and Department of Chemistry, University of Houston, Houston, Texas 77204-5003, United States}

\author{Bernd Lorenz}
\affiliation{Texas Center for Superconductivity, and Department of Physics, University of Houston, Houston, Texas 77204-5005, United States}

\author{Yurii Skourski}
\affiliation{Dresden High Magnetic Field Laboratory, Helmholtz-Zentrum Dresden-Rossendorf, 01314 Dresden, Germany}

\author{Helge Rosner}
\affiliation{Max Planck Institute for Chemical Physics of Solids, N\"{o}thnitzer Str. 40, 01187 Dresden, Germany}


\begin{abstract}
Based on density functional theory band structure calculations, quantum Monte-Carlo simulations, and high-field magnetization measurements, we address the microscopic magnetic model of BaAg$_2$Cu[VO$_4]_2$ that was recently proposed as a spin-$\frac12$ anisotropic triangular lattice system. We show that the actual physics of this compound is determined by a peculiar superposition of ferromagnetic and antiferromagnetic uniform spin chains with nearest-neighbor exchange couplings of $J_a^{(1)}\simeq -19$~K and $J_a^{(2)}\simeq 9.5$~K, respectively. The two chains featuring different types of the magnetic exchange perfectly mimic the specific heat of a triangular spin lattice, while leaving a clear imprint on the magnetization curve that is incompatible with the triangular-lattice model. Both ferromagnetic and antiferromagnetic spin chains run along the crystallographic $a$ direction, and slightly differ in the mutual arrangement of the magnetic CuO$_4$ plaquettes and non-magnetic VO$_4$ tetrahedra. These subtle structural details are, therefore, crucial for the ferromagnetic or antiferromagnetic nature of the exchange couplings, and put forward the importance of comprehensive microscopic modeling for a proper understanding of quantum spin systems in transition-metal compounds.
\end{abstract}

\maketitle

\section{Introduction}
Frustration and dimensionality are two crucial parameters underlying the physics of magnetic systems. In insulators, these parameters rarely correlate with the apparent features of the atomic arrangement, because superexchange couplings are highly sensitive to details of the electronic structure and to positions of non-magnetic atoms linking the magnetic sites. While computational techniques based on electronic structure calculations developed into a powerful tool for elucidating spin lattices of complex materials, simple phenomenological criteria are equally important for the preliminary assessment of the experimental data and the compound under consideration.

The best-known phenomenological criterion of the magnetic frustration is the $|\theta|/T_N$ ratio. It compares the Curie-Weiss temperature $\theta$, which is often considered as an effective energy scale of the magnetic couplings, to the magnetic ordering temperature $T_N$.\cite{ramirez1994,*greedan2001} High $|\theta|/T_N$ ratios are believed to indicate strong frustration, although this rule will only hold for simple systems with few exchange couplings and well-established dimensionality. Thus, the $|\theta|/T_N\simeq 50-100$ ratio is easily obtained even in non-frustrated quasi-one-dimensional (1D) systems, where strong quantum fluctuations due to the weak interchain couplings effectively prevent the system from long-range ordering down to low temperatures.\cite{yasuda2005,kojima1997,*rosner1997,lancaster2006,belik2005,*johannes2006} Another possible scenario is that of magnets with strong dimer correlations, where the long-range-ordered state competes with the disordered singlet ground state, and the ordering temperature $T_N$ may be strongly reduced without any frustration involved.\cite{troyer1997,kadono1996,*matsumoto1996,katoh1994,*wang1997} The low $|\theta|/T_N$ ratio can be equally deceptive, because $\theta$ is in fact a linear combination of different exchange couplings that can be much smaller than the effective energy scale of the system. For example, the coexistence of ferromagnetic (FM) and antiferromagnetic (AFM) couplings renders $\theta$ and $|\theta|/T_N$ low even in strongly frustrated magnets.\cite{nath2008a,nath2008b}

The phenomenological assessment of the frustration in a magnetic system has to be backed by additional criteria. Magnetic specific heat is an especially appealing quantity, because it is expressed in absolute units and does not require an ambiguous reference to the effective energy scale of the system. Further, the magnetic specific heat distinguishes between the effects of dimensionality and frustration, with the latter leading to a much stronger reduction in the maximum of the magnetic specific heat ($C_m$). For example, the spin-$\frac12$ square lattice (two-dimensional, non-frustrated) reveals the maximum of $C_m/R\simeq 0.44$, the spin-$\frac12$ uniform chain (one-dimensional, non-frustrated) shows a lower maximum of $C_m/R\simeq 0.35$, but the specific heat maximum for the spin-$\frac12$ triangular lattice (two-dimensional, frustrated) is even lower, $C_m/R\simeq 0.22$.\cite{bernu2001} The reduced magnetic specific heat is a seemingly unambiguous measure of the frustration. It can be equally used to identify strongly frustrated spin systems\cite{nath2008a,nath2008b,okamoto2007,*lawler2008} or to refute premature conclusions on the strong frustration.\cite{kaul2005,*rosner2002} However, this phenomenological criterion is not universal, as we demonstrate below.

\begin{figure}
\includegraphics{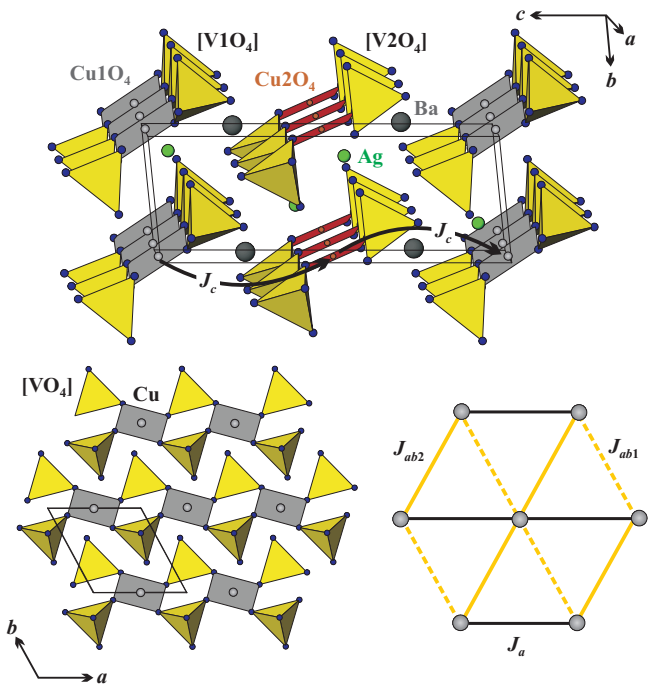}
\caption{\label{fig:str}
(Color online) Top panel: perspective view of the BaAg$_2$Cu[VO$_4]_2$ structure showing alternating layers consisting of VO$_4$-bridged chains of Cu1 and Cu2, respectively. Different colors (shadings) identify the inequivalent CuO$_4$ plaquettes and their slightly different orientation. $J_c$ refers to the interlayer coupling. Bottom panel: a single layer in the $ab$ plane (left) and the respective spin lattice with the intrachain coupling $J_a$ as well as interchain couplings $J_{ab1}$ and $J_{ab2}$ (right).
}
\end{figure}
In our study, the breakdown of the simple relationship between the magnetic specific heat and the frustration is related to a peculiar superexchange scenario in BaAg$_2$Cu[VO$_4]_2$. This compound has a fairly complex crystal structure with magnetic Cu$^{2+}$ ions interspersed between the non-magnetic [VO$_4]^{3-}$ tetrahedra as well as Ba$^{2+}$ and Ag$^{+}$ cations.\cite{moeller2011} The spatial arrangement of Cu$^{2+}$ (Fig.~\ref{fig:str}) resembles a weakly anisotropic triangular lattice with the intraplane Cu--Cu distances of 5.45~\r A ($J_a$), 5.63~\r A ($J_{ab1}$), and 5.69~\r A ($J_{ab2}$) and the interplane distance of 7.20~\r A ($J_c$). This lattice topology should induce magnetic frustration, as further corroborated by the magnetic specific heat that reaches the maximum value of $C_m/R\simeq 0.22$ and strongly resembles theoretical predictions for the spin-$\frac12$ triangular lattice.\cite{moeller2011}

In the following, we will show that the reduced $C_m$ has a different origin, and arises from a peculiar superposition of FM and AFM spin chains. The system is, therefore, quasi-one-dimensional and only weakly frustrated, in contrast to the straight-forward phenomenological assessment. To support the one-dimensional scenario, we perform extensive band structure calculations combined with the fitting of magnetization and specific heat data. We also present original experimental results on the high-field magnetization that unequivocally rules out the triangular-lattice spin model for BaAg$_2$Cu[VO$_4]_2$.

\section{Methods}

Our microscopic magnetic model of BaAg$_2$Cu[VO$_4]_2$ is based on full-potential scalar-relativistic density functional theory (DFT) band structure calculations performed in the \texttt{FPLO} code\cite{fplo} implementing the basis set of local orbitals. We used the local density approximation (LDA) with the Perdew-Wang parametrization for the exchange-correlation potential.\cite{pw92} The $k$ meshes of 292~points and 64~points in the symmetry-irreducible part of the first Brillouin zone were chosen for the crystallographic unit cell and supercell, respectively. Correlation effects were treated on a model level or within the mean-field local spin-density approximation (LSDA)+$U$ approach, as further described in Sec.~\ref{sec:dft}.

Thermodynamic properties were calculated with the \texttt{loop}\cite{loop} and \verb|dirloop_sse| (directed loop in stochastic series expansion representation)\cite{dirloop} quantum Monte-Carlo (QMC) algorithms implemented in the \texttt{ALPS} simulation package.\cite{alps} Simulations were done for finite lattices with periodic boundary conditions. We used two independent chains containing $L=40$ sites each. This chain length is sufficient to eliminate finite-size effects for thermodynamic properties within the temperature range under investigation.

Powder samples of BaAg$_2$Cu[VO$_4]_2$ were prepared according to the method described in Ref.~\onlinecite{moeller2011}. Magnetic susceptibility was measured with MPMS SQUID magnetometer in the temperature range $2-380$~K in the applied field of 0.1~T. Magnetization isotherm was collected at 1.5~K using the pulsed magnet installed in Dresden High Magnetic Field Laboratory. Details of the experimental procedure are described elsewhere.\cite{tsirlin2009} The low-temperature heat capacity was measured above 0.5~K by a relaxation method using the $^3$He option of the Physical Property Measurement System (PPMS, Quantum Design).

\section{Microscopic magnetic model}
\label{sec:dft}
\begin{figure}
\includegraphics{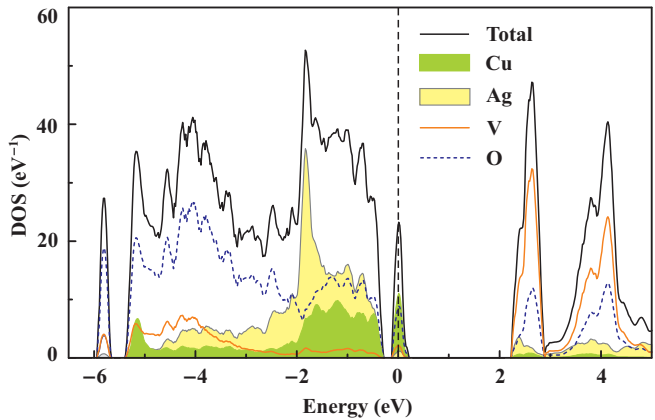}
\caption{\label{fig:dos}
(Color online) LDA density of states for BaAg$_2$Cu[VO$_4]_2$. The Fermi level is at zero energy.
}
\end{figure}

LDA results for the band structure of BaAg$_2$Cu[VO$_4]_2$ (Fig.~\ref{fig:dos}) closely follow expectations for a Cu$^{2+}$-based insulating compound.\cite{cu2v2o7,mazurenko2007,tsirlin2010b,tsirlin2011,*janson2011} Oxygen $2p$ states between $-6$ and $-2$~eV are surmounted by Ag $4d$ and Cu $3d$ bands. The states above 2~eV originate from unfilled V $3d$ orbitals. While silver states are mostly found below $-0.3$~eV, Cu $3d$ states additionally form narrow bands in the vicinity of the Fermi level. The calculated partial densities of states confirm the anticipated valences of Ag$^{1+}$ ($4d^{10}$), Cu$^{2+}$ ($3d^9$), and V$^{5+}$ ($3d^0$), and identify Cu$^{2+}$ ions as the magnetic sites in the structure. The metallic LDA energy spectrum violates the insulating nature of the compound, as evidenced by the dark-yellow color of BaAg$_2$Cu[VO$_4]_2$. This discrepancy is well understood, given the importance of correlation effects for the partially filled Cu $3d$ shell and the severe underestimation of such correlations in LDA. The missing correlations can be introduced on the model level, or by a mean-field LSDA+$U$ procedure.

\begin{figure}
\includegraphics{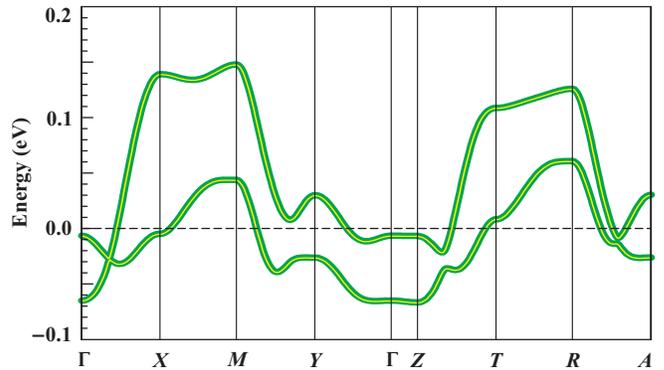}
\caption{\label{fig:band}
(Color online) LDA bands (thin light lines) and the fit with the tight-binding model (thick dark lines). The $k$ path is defined as follows: $\Gamma(0,0,0)$, $X(0.5,0,0)$, $M(0.5,0.5,0)$, $Y(0,0.5,0)$, $Z(0,0,0.5)$, $T(0.5,0,0.5)$, $R(0.5,0.5,0.5)$, and $A(0,0.5,0.5)$, where the coordinates are given in units of the respective reciprocal lattice parameters.
}
\end{figure}

Following the first approach to the treatment of correlations, we consider in more detail the narrow bands in the vicinity of the Fermi level (Fig.~\ref{fig:band}). The two bands can be assigned to two inequivalent Cu sites in the crystal structure. Both bands have the $d_{x^2-y^2}$ orbital character, with $x$ and $y$ axes directed along shorter Cu--O bonds. In BaAg$_2$Cu[VO$_4]_2$, the local environment of Cu$^{2+}$ resembles a severely elongated octahedron, CuO$_{4+2}$, with four short Cu--O bonds ($1.96-1.97$~\r A) lying in the plane and two long bonds (2.44~\r A) perpendicular to this plane. Therefore, the $d_{x^2-y^2}$ orbital is the highest-lying crystal-field level in agreement with the LDA results.

To fit the $d_{x^2-y^2}$ bands with the tight-binding model, we construct Wannier functions localized on Cu sites.\cite{wannier} The fit perfectly reproduces the calculated band structure (Fig.~\ref{fig:band}), and yields Cu--Cu hopping parameters $t_i$ (Table~\ref{tab:exchange}). Mapping the tight-binding model onto a one-orbital Hubbard model with the effective on-site Coulomb repulsion $U_{\eff}=4.5$~eV,\cite{cu2v2o7,tsirlin2011,*janson2011} we find the anticipated strongly correlated regime ($t_i\ll U_{\eff}$), and utilize second-order perturbation theory for analyzing the lowest-lying (magnetic) excitations. This way, AFM contributions to the exchange couplings are evaluated as $J_i^{\AFM}=4t_i^2/U_{\eff}$.

The results of our model analysis are summarized in Table~\ref{tab:exchange}.\footnote{%
Further terms $t_i$ are below 5~meV and can be neglected in the present analysis.} While AFM couplings in BaAg$_2$Cu[VO$_4]_2$ are mostly weak, we find the sizable AFM coupling $J_a^{(2)}$ along the $a$ direction. Remarkably, this AFM coupling along $a$ (denoted $J_a$) is observed for the Cu2 site and not for the Cu1 site, as emphasized by the superscripts $(1)$ and $(2)$ in the notation of $J_i$. This observation puts forward one important feature of the BaAg$_2$Cu[VO$_4]_2$ structure. The two Cu sites in BaAg$_2$Cu[VO$_4]_2$ are very similar and look nearly identical with respect to the geometry of individual superexchange pathways (Table~\ref{tab:distances}). The Cu1--Cu1 and Cu2--Cu2 distances in the $ab$ plane are equal because of the constraints imposed by the lattice translations. However, our microscopic analysis puts forward important differences between the deceptively similar superexchange pathways within the Cu1 and Cu2 sublattices  (Table~\ref{tab:exchange}). This difference gives a clue to understand the magnetism of BaAg$_2$Cu[VO$_4]_2$, and will be discussed in more detail below.

\begin{table}
\caption{\label{tab:exchange}
Cu--Cu distances (in~\r A), hoppings $t_i$ (in~meV), and exchange couplings $J_i$ (in~K) in BaAg$_2$Cu[VO$_4]_2$. The AFM contributions $J_i^{\AFM}$ are calculated as $4t_i^2/U_{\eff}$ with $U_{\eff}=4.5$~eV; the full exchange couplings $J_i$ are obtained from LSDA+$U$ calculations ($U_d=6$~eV, $J_d=1$~eV); and $J_i^{\FM}=J_i-J_i^{\AFM}$. The notation of $J_i$ is illustrated in Fig.~\ref{fig:str}.
}
\begin{ruledtabular}
\begin{tabular}{rcr@{\hspace{1em}}rrr}
                 & Distance & $t_i$ & $J_i^{\AFM}$ & $J_i^{\FM}$ & $J_i$  \\\hline
 $J_a^{(1)}$     &  5.45    & $-11$ &     1        &   $-21$     & $-20$  \\\smallskip
 $J_a^{(2)}$     &  5.45    & $-43$ &    19        &   $-16$     &    3   \\\smallskip
 $J_{ab1}^{(1)}$ &  5.63    & $-8$  &     1        &   $-3$      &  $-2$  \\\smallskip
 $J_{ab1}^{(2)}$ &  5.63    &    0  &     0        &     0       &    0   \\\smallskip
 $J_{ab2}^{(1)}$ &  5.69    &    0  &     0        &   $-0.3$    & $-0.3$ \\\smallskip
 $J_{ab2}^{(2)}$ &  5.69    &    0  &     0        &   $-0.3$    & $-0.3$ \\
 $J_c$           &  7.20    &   11  &     1        &   $-0.3$    &   0.7  \\
\end{tabular}
\end{ruledtabular}
\end{table}
The FM part of the superexchange originates from processes beyond the one-orbital model employed in our tight-binding analysis. In cuprates, FM interactions are generally ascribed to the Hund's coupling on the ligand site\cite{mazurenko2007} and can be evaluated by mapping total energies for different collinear spin configurations onto the classical Heisenberg model. The total energies are obtained from spin-polarized band structure calculations with LSDA+$U$ as the mean-field correction for correlation effects. Following previous studies of Cu$^{2+}$-based compounds,\cite{cu2v2o7,janson2011} we use the around-mean-field double-counting correction scheme, the on-site Coulomb repulsion parameter $U_d=6$~eV, and the Hund's exchange parameter $J_d=1$~eV. In the case of BaAg$_2$Cu[VO$_4]_2$, alterations of $U_d$ and the double-counting correction scheme have marginal influence on the results, and do not change the qualitative microscopic scenario.

\begin{figure}
\includegraphics{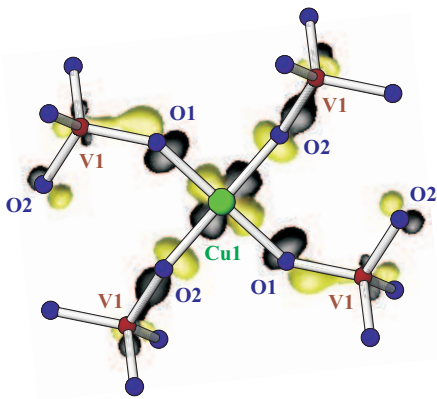}
\caption{\label{fig:wannier}
(Color online) Wannier function based on the Cu $d_{x^2-y^2}$ orbital.
}
\end{figure}
The total exchange couplings $J_i$ based on the LSDA+$U$ calculations are listed in the last column of Table~\ref{tab:exchange}. We find comparable FM contributions to the couplings $J_a^{(1)}$ and $J_a^{(2)}$ along the $a$ direction. Owing to the larger AFM contribution to $J_a^{(2)}$, this coupling remains weakly AFM, while $J_a^{(1)}$ becomes FM. Other couplings show small FM contributions and hover around zero. The LSDA+$U$ calculations confirm the leading couplings along $a$ as well as the notable difference between $J_a^{(1)}$ and $J_a^{(2)}$. Before comparing our magnetic model to the experimental data, we further comment on the microscopic origin of different exchange couplings in the Cu1 and Cu2 sublattices of BaAg$_2$Cu[(VO$_4]_2$.

The sizable FM and AFM contributions are identified for the exchange couplings $J_a^{(1)}$ and $J_a^{(2)}$, only. This finding is easily rationalized based on the magnetic $d_{x^2-y^2}$ orbital of the Cu$^{2+}$ ions. The crystal structure is best viewed in terms of the CuO$_4$ plaquettes entailing the magnetic orbitals. This representation underscores the 1D nature of the structure (Fig.~\ref{fig:str}), and illustrates the quasi-1D magnetic behavior. However, unlike the well-known spin-chain Cu$^{2+}$ compounds, such as Sr$_2$CuO$_3$ (Ref.~\onlinecite{kojima1997,*rosner1997}) and CuPzN,\cite{lancaster2006,hammar1999,*stone2003} BaAg$_2$Cu[VO$_4]_2$ reveals a combination of two inequivalent spin chains with strikingly different exchange couplings.

\begin{figure}
\includegraphics{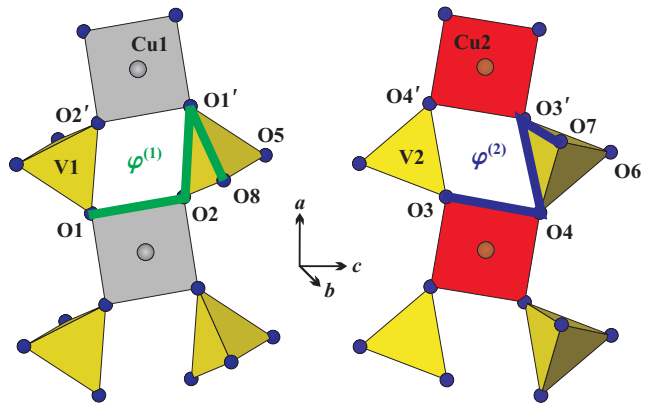}
\caption{\label{fig:geometry}
(Color online) Comparison of the Cu1 (left) and Cu2 (right) chains in the BaAg$_2$Cu[VO$_4]_2$ structure. Note the different orientations of the VO$_4$ tetrahedra with respect to the CuO$_4$ plaquettes, as quantified by the respective dihedral angles $\varphi^{(1)}$ and $\varphi^{(2)}$.}
\end{figure}
According to Table~\ref{tab:exchange}, both $J_a^{(1)}$ and $J_a^{(2)}$ feature similar FM contributions, yet very different AFM exchanges arising from different Cu--Cu hoppings in the effective one-orbital model. To elucidate the origin of these couplings, we consider the Wannier functions localized on Cu sites. Apart from the Cu $d_{x^2-y^2}$ orbital forming the core of the Wannier function, we find sizable contributions from oxygen $2p$ and vanadium $3d$ orbitals (Fig.~\ref{fig:wannier}). These contributions can also be observed in the LDA energy spectrum (Fig.~\ref{fig:dos}). The Wannier functions of the neighboring Cu atoms overlap on the vanadium sites, where each Wannier function features a different $3d$ orbital of vanadium. This leads to the Hund's exchange on the vanadium site and explains the sizable FM contributions to $J_a^{(1)}$ and $J_a^{(2)}$, in contrast to the very low FM contributions to other nearest-neighbor couplings having similar Cu--Cu distances (Table~\ref{tab:exchange}). Note that a comparable $J^{\FM}\simeq -15$~K has been found in $\beta$-Cu$_2$V$_2$O$_7$, where vanadium $3d$ orbitals also contribute to the Cu-based Wannier functions.\cite{cu2v2o7}

\begin{table}
\caption{\label{tab:distances}
Interatomic distances (in~\r A) and angles (in~deg) in the BaAg$_2$Cu[VO$_4]_2$ structure. The columns refer to the Cu1 and Cu2 layers, as shown in Fig.~\ref{fig:str}. The notation of individual atoms follows Fig.~\ref{fig:geometry} (see text for details).
}
\begin{ruledtabular}
\begin{tabular}{cr@{\hspace{3em}}cr}
  Cu1--O1 & $2\times 1.973$ & Cu2--O3 & $2\times 1.969$ \\
  Cu1--O2 & $2\times 1.974$ & Cu2--O4 & $2\times 1.959$ \\
  Cu1--O8 & $2\times 2.436$ & Cu2--O7 & $2\times 2.444$ \\
  V1--O1  & 1.749           & V2--O3  & 1.757           \\
  V1--O2  & 1.740           & V2--O4  & 1.755           \\
  V1--O5  & 1.681           & V2--O6  & 1.674           \\
  V1--O8  & 1.713           & V2--O7  & 1.713           \\
  O1--O2  & 2.884           & O3--O4  & 2.907           \\\hline
  Cu1--O1$'$--O2       & 113.1 & Cu2--O4$'$--O3    & 113.4 \\
  Cu1--O2$'$--O1       & 145.1 & Cu2--O3$'$--O4    & 144.6 \\
  $\varphi^{(1)}$      & 123.7 & $\varphi^{(2)}$   & 102.2 \\
\end{tabular}
\end{ruledtabular}
\end{table}
We now consider different AFM contributions to $J_a^{(1)}$ and $J_a^{(2)}$. Geometrical parameters summarized in Table~\ref{tab:distances} demonstrate a striking similarity between the respective superexchange pathways for Cu1 and Cu2. The only notable difference is the orientation of the VO$_4$ tetrahedra with respect to the chains. Naively, the position of the tetrahedra is described by the O8--O2--O1 and O7--O4--O3 angles (Fig.~\ref{fig:geometry}). However, these do not account for the different tilting of the Cu1O$_4$ and Cu2O$_4$ plaquettes with respect to the $a$ axis (Fig.~\ref{fig:str}). Therefore, we use dihedral angles $\varphi$ referring to the O1$'$--O8--O2 and O1--O2--O1$'$--O2$'$ planes for Cu1 ($\varphi^{(1)}$), and to the O3$'$--O7--O4 and O3--O4--O3$'$--O4$'$ planes for Cu2 ($\varphi^{(2)}$). According to Table~\ref{tab:distances}, the difference between $\varphi^{(1)}$ and $\varphi^{(2)}$ is as large as 21.5~deg, thus to be considered as the main feature to account for the different Cu--Cu hoppings $t_a^{(1)}$ and $t_a^{(2)}$.

To explore the role of the dihedral angles $\varphi$, we construct fictitious model structures with the VO$_4$ tetrahedra rotated about the O--O edges (O1--O2 and \mbox{O3--O4} for V1 and V2, respectively). This way, we are able to tune $\varphi^{(1)}$ toward $\varphi^{(2)}=102.2$~deg and enhance $t_a^{(1)}$ to 21~meV (compare to $-11$~meV at the experimental $\varphi^{(1)}=123.7$~deg), or change $\varphi^{(2)}$ toward $\varphi^{(1)}=123.7$~deg, thus reducing $t_a^{(2)}$ to 3~meV (compare to $-43$~meV at the experimental $\varphi^{(1)}=102.2$~deg). Overall, a change of orientation by approximately 22~deg is accompanied by a $\Delta(t_a)$ of 32~meV and 46~meV, respectively. Therefore, the orientation of the non-magnetic VO$_4$ tetrahedra is of crucial importance for the Cu--Cu hoppings and AFM superexchange. Note, however, that this geometrical parameter is not unique, and the specific arrangement of the CuO$_4$ plaquettes with respect to the chain direction (Fig.~\ref{fig:str} and Fig.~\ref{fig:geometry}) is also responsible for the large AFM contribution to $J_a^{(2)}$.

\section{Experimental data}
\label{sec:exp}
\begin{figure}
\includegraphics{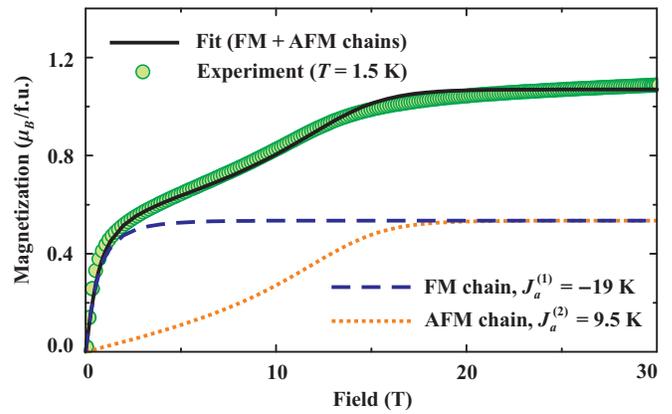}
\caption{\label{fig:mvsh}
(Color online) Magnetization isotherm of BaAg$_2$Cu[VO$_4]_2$ measured at 1.5~K (filled circles) and the fit with a combination of FM and AFM spin chains (solid line). The contributions of the FM (Cu1) and AFM (Cu2) chains are shown by the dashed and dotted lines, respectively.
}
\end{figure}

The DFT results summarized in Table~\ref{tab:exchange} identify the spin lattice of BaAg$_2$Cu[VO$_4]_2$ as a system of weakly interacting inequivalent spin chains with the intrachain couplings $J_a^{(1)}$ and $J_a^{(2)}$, respectively. While $J_a^{(1)}$ is clearly FM, $J_a^{(2)}$ is weakly AFM and probably close to zero. This qualitative scenario is verified by the magnetization isotherm measured at 1.5~K. Previous measurements\cite{moeller2011} in fields up to 5~T showed that half of the Cu spins seem to saturate around 1.5~T. Here we extend our study into the behavior of the magnetization in higher fields (Fig.~\ref{fig:mvsh}). Based on these high-field measurements, we show that the magnetization of BaAg$_2$Cu[VO$_4]_2$ is further increased between 1.5~T and 16~T, where the full saturation with $M\simeq 1.08$~$\mu_B$/f.u. is reached. This peculiar behavior apparently contradicts the conjecture on the triangular spin lattice that would lead to a smooth increase in the magnetization between zero field and the saturation field.\cite{[{For example: }][{}]tokiwa2006,*starykh2007}

The experimental magnetization curve is readily elucidated by our microscopic model. While half of the spins comprising the FM spin chains (Cu1) align with the field already at $1.0-1.5$~T once thermal fluctuations are suppressed, the remaining spins (Cu2) are coupled antiferromagnetically and require larger fields to overcome the AFM interactions. This behavior strongly reminds of a two-sublattice ferrimagnet, where half of the maximum magnetization is recovered in low fields, while larger fields are required to flip one of the sublattices. Note, however, that BaAg$_2$Cu[VO$_4]_2$ is not in a magnetically ordered state at 1.5~K, hence no magnetization hysteresis is observed. The long-range magnetic order in BaAg$_2$Cu[VO$_4]_2$ is established below $T_C\simeq 0.7$~K and is further discussed in Sec.~\ref{sec:discussion}.

\begin{figure}
\includegraphics{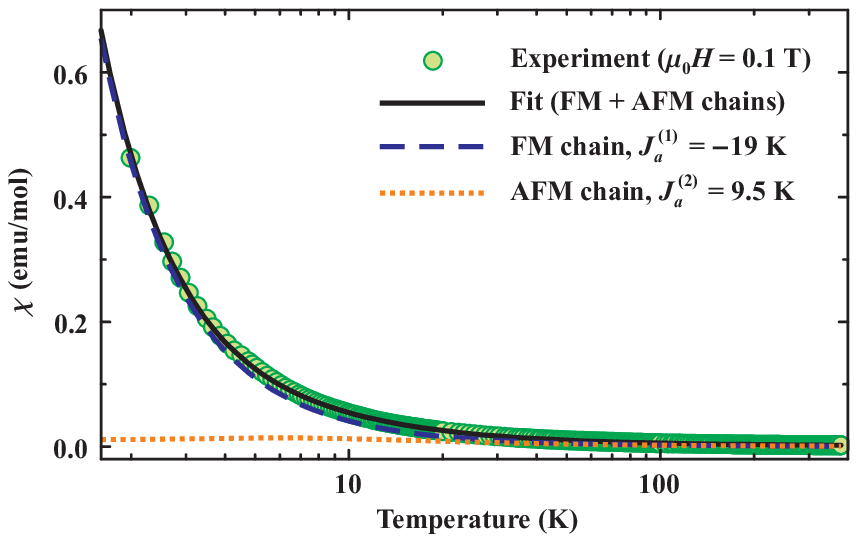}
\caption{\label{fig:chi}
(Color online) Magnetic susceptibility of BaAg$_2$Cu[VO$_4]_2$ measured in the applied field of 0.1~T (filled circles) and the fit with a combination of FM and AFM spin chains (solid line). The contributions of the FM (Cu1) and AFM (Cu2) chains are shown by the dashed and dotted lines, respectively.
}
\end{figure}
The above qualitative picture can be quantified by fitting the experimental magnetization data.\footnote{%
We simulate the magnetization isotherm at 2.5~K, which is somewhat larger than the experimental temperature of 1.5~K. The larger temperature used in the simulation is a necessary compromise between the sharp saturation of the FM (Cu1) subsystem around 2~T and the very broad saturation anomaly around 14~T. This difference between the two saturation processes is likely a signature of a spurious heating driven by a magnetocaloric effect in the pulsed-field experiment. Further measurements in static fields would be helpful for getting more accurate magnetization data.} In BaAg$_2$Cu[VO$_4]_2$, field dependence of the magnetization (Fig.~\ref{fig:mvsh}) and temperature dependence of the susceptibility (Fig.~\ref{fig:chi}) are complimentary. The magnetization isotherm is sensitive to the AFM exchange $J_a^{(2)}$ that determines the saturation field, but the alignment of the FM component mostly depends on thermal fluctuations so that $J_a^{(1)}$ cannot be determined precisely. In contrast, the FM chains coupled by $J_a^{(1)}$ produce the dominant contribution to the susceptibility,\footnote{%
The effect of $J_a^{(2)}$ is still visible in the Curie-Weiss temperature $\theta\simeq -3.3$~K, which is much lower that $J_a^{(1)}/2\simeq -9.5$~K expected for a single FM spin chain.} which gives an accurate estimate for $J_a^{(1)}$, while leaving certain ambiguity for $J_a^{(2)}$. The two sets of data are successfully fitted with the same model parameters: $J_a^{(1)}\simeq -19$~K, $J_a^{(2)}\simeq 9.5$~K, $g\simeq 2.16$ (Figs.~\ref{fig:mvsh} and~\ref{fig:chi}). We also included a temperature-independent contribution to the susceptibility $\chi_0\simeq 9\times 10^{-4}$~emu/mol that accounts for the van Vleck paramagnetism and core diamagnetism. Our fitted $g$-value is in excellent agreement with the experimental powder-averaged $\bar g=2.18$.\cite{moeller2011} While $J_a^{(1)}$ closely follows the DFT prediction (Table~\ref{tab:exchange}), the computational estimate of $J_a^{(2)}$ is less accurate, although still acceptable considering the low energy scale of the exchange couplings in BaAg$_2$Cu[VO$_4]_2$.\footnote{%
The supercell procedure evaluates the exchange couplings $J_i$ as the difference in total energies ($E_i$). In BaAg$_2$Cu[VO$_4]_2$, the typical $J_i/E_i$ ratio is below $10^{-9}$.}

\begin{figure}
\includegraphics{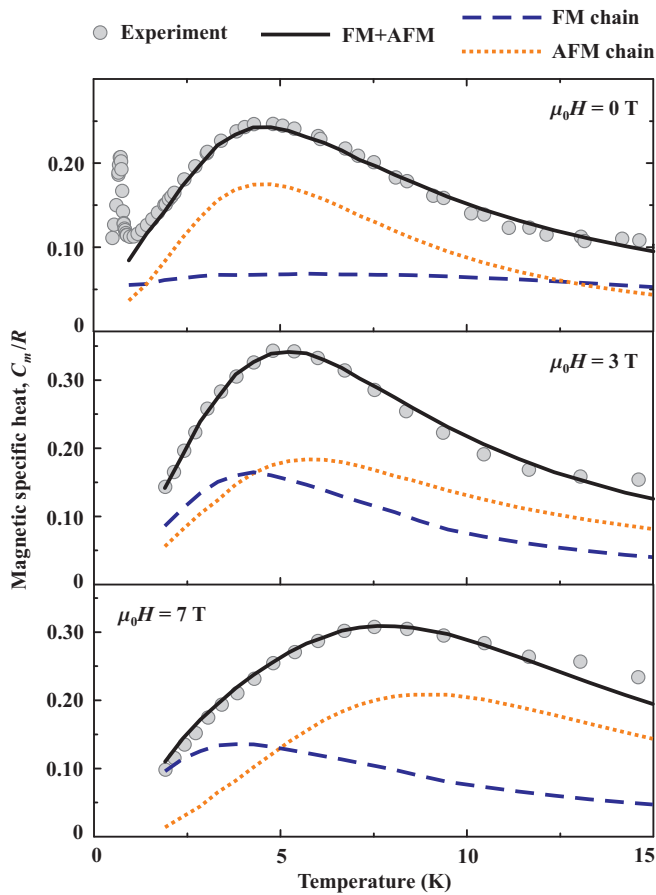}
\caption{\label{fig:heat}
(Color online) Magnetic part of the specific heat $[C_m(T)]$ divided by the gas constant $(R)$ for BaAg$_2$Cu[VO$_4]_2$ measured in zero field (top) and in applied fields of 3~T (middle) and 7~T (bottom). The simulated curves for the combination of FM and AFM spin chains are shown by solid lines, whereas the dashed and dotted lines denote the contributions of the FM (Cu1) and AFM (Cu2) spin chains, respectively. Experimental data (circles) are taken from Ref.~\onlinecite{moeller2011}. The model parameters $J_a^{(1)}=-19$~K and $J_a^{(2)}=9.5$~K are extracted from the fits to the magnetization data (Figs.~\ref{fig:mvsh} and~\ref{fig:chi}). Therefore, we compare our model to the experiment with no adjustable parameters.
}
\end{figure}
Figures~\ref{fig:mvsh} and~\ref{fig:chi} illustrate the contributions of the FM and AFM components to the magnetization and susceptibility of BaAg$_2$Cu[VO$_4]_2$, respectively. The FM chains lead to the sharp increase in the susceptibility at low temperatures, while the contribution of the AFM chains is barely visible on the same scale. The contribution of the FM chains to the magnetization isotherm is saturated at low fields and corresponds to one half of the maximum magnetization, because half of the Cu atoms belong to the FM chains. The magnetization of the AFM chains is linear at low fields, bends upward above 7~T, and finally saturates around 16~T where the full alignment of spins is achieved.

We will now test our quasi-1D model against the experimental specific heat data showing the strongly reduced maximum that might be characteristic of a spin-$\frac12$ triangular lattice. We use the fitted parameters based on the magnetization data and, therefore, compare our model to the experiment with no adjustable parameters.\footnote{%
Since Ref.~\onlinecite{moeller2011} reports an about 10~\% underestimate in the magnetic entropy, we increase the experimental magnetic specific heat by 10~\%.} Fig.~\ref{fig:heat} presents the magnetic specific heat data measured in zero field and in two representative applied fields along with the simulated curves. The remarkable agreement between the experiment and the model prediction confirms our microscopic scenario, and suggests that the strongly reduced specific heat maximum, especially in zero field, is not an unambiguous footprint of the magnetic frustration. 

In zero field, the specific heat maximum closely follows the contribution of the AFM spin chains, while the FM chains with the stronger coupling $J_a^{(1)}\simeq -19$~K provide a temperature-independent ``background'' below 15~K. The applied field of 3~T increases the maximum up to $C_m/R\simeq 0.32$. The stronger field of 7~T additionally shifts the maximum to higher temperatures. Both effects are perfectly reproduced by our microscopic model. Magnetic fields transform the temperature-independent zero-field specific heat of the FM chains into a small maximum at $3.5-4.0$~K. This maximum of the FM contribution weakly depends on the field, because the FM (Cu1) subsystem is saturated above 2~T (Fig.~\ref{fig:mvsh}). By contrast, the contribution of the AFM chains shows a pronounced field dependence that underlies the evolution of the experimental magnetic specific heat in the applied field.

\section{Discussion and summary}
\label{sec:discussion}
The combination of DFT calculations and QMC fits to the experimental data gives compelling evidence for the quasi-1D magnetic behavior of BaAg$_2$Cu[VO$_4]_2$. The superposition of FM and AFM spin chains with different magnitudes of the exchange couplings results in peculiar and perplexing thermodynamic properties. While the zero-field specific heat resembles the typical response of the spin-$\frac12$ triangular lattice, the magnetization isotherm is reminiscent of a system with two different magnetic sublattices, and underpins the proposed magnetic model.

Based on our microscopic analysis, we establish the spin lattice of BaAg$_2$Cu[VO$_4]_2$ as a peculiar derivative of conventional Heisenberg spin chains with nearest-neighbor exchange coupling $J$. This model was widely studied for both FM and AFM $J$,\cite{bethe1931,*takahashi1971,bonner1964,*griffiths1964,*parkinson1985,kluemper2000,*johnston2000} but the combination of FM and AFM spin chains was neither considered theoretically nor encountered experimentally. 

The superposition of inequivalent spin chains is a challenge for ``non-local'' experimental techniques, such as thermodynamic measurements or inelastic neutron scattering, that probe the system as a whole. These methods inevitably blend the signals of different sublattices, and generally lead to a complex response that can be fully elucidated based on the microscopic approach only. A more direct experimental information could be extracted from ``local'' methods, which probe different magnetic sublattices independently. For example, an elegant way to study the physics of BaAg$_2$Cu[VO$_4]_2$ further could be nuclear magnetic resonance (NMR) on $^{51}$V nuclei. The inequivalent vanadium sites V1 and V2 are coupled to Cu1 and Cu2, respectively. Owing to the very similar local environment, the signals from these two vanadium sites should perfectly overlap at high temperatures. At low temperatures though, the lines will split because of the different Knight shifts resulting from the disparate local magnetization in the vicinity of the FM and AFM spin chains. Therefore, the NMR experiment should be a valuable additional experimental test of our microscopic model.

\begin{figure}
\includegraphics{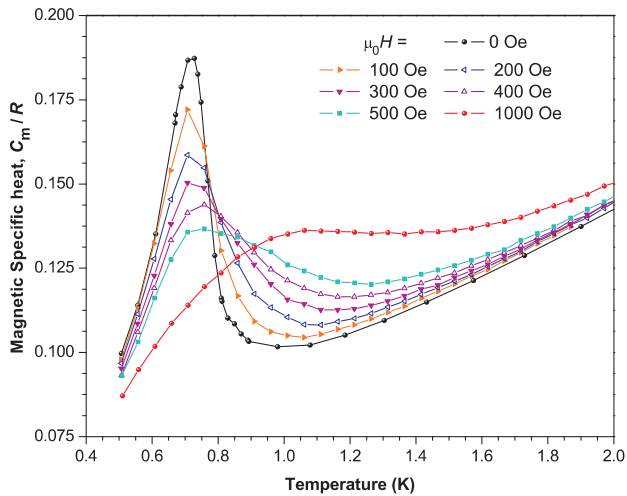}
\caption{\label{fig:Tc}
(Color online) Field-dependent magnetic part of the specific heat, $C_m(T)$, divided by $R$ for BaAg$_2$Cu[VO$_4]_2$ at low temperatures, showing the behavior typical for a ferromagnetic or ferrimagnetic transition.
}
\end{figure}
Another interesting problem is the long-range-ordered (LRO) ground state of BaAg$_2$Cu[VO$_4]_2$. While isolated spin chains do not show the LRO even at zero temperature, interchain couplings induce the LRO state at a finite temperature,\cite{schulz1996,sandvik1999,irkhin2000} irrespective of the weak frustration that could be induced by the triangular arrangement.\cite{bocquet2001,*bocquet2002} In BaAg$_2$Cu[VO$_4]_2$, specific-heat measurements reveal the sharp anomaly at $T_C\simeq 0.7$~K in zero field. This anomaly is drastically suppressed even in weak magnetic fields (Fig.~\ref{fig:Tc}, see also Ref.~\onlinecite{moeller2011}), as typical for a ferromagnetic transition, or -- more generally -- for an LRO state with non-zero net magnetization. Such ground state can be indeed derived from our microscopic model and explained in terms of a two-sublayer system with the interlayer exchange coupling $J_c$. As outlined above, these sublayers are stacked along the $c$-axis in an alternate fashion. Each plane consists either of FM (Cu1) or AFM (Cu2) spin chains, respectively (Fig.~\ref{fig:str}).

The Cu1 spins within the FM spin chains prefer the parallel alignment so that a FM sublattice is formed. The Cu2 spins are expected to be ordered antiferromagnetically along $a$ and form an AFM sublattice. The nature of the interchain couplings is more difficult to establish because of their lower energy scale that might allow for additional, non-isotropic contributions, such as dipolar interactions. However, even the isotropic (Heisenberg) model based on DFT enables us to make a plausible conjecture about the ground state. The FM couplings $J_{ab1}$ and $J_{ab2}$ in the $ab$ plane are compatible with both FM and AFM exchange along $a$. These couplings should reinforce the formation of the FM sublattice for Cu1 and the AFM sublattice for Cu2. The AFM coupling $J_c$ along $c$ introduces a weak frustration, but its effect should be small. In summary, BaAg$_2$Cu[VO$_4]_2$ entails two inequivalent sublattices and presents a peculiar example of a spin-$\frac12$ system with non-zero net magnetization. 

From phenomenological point of view, a similar ground state with the non-zero net magnetization has been recently observed in the spin-$\frac12$ ferrimagnet Cu$_2$OSeO$_3$.\cite{bos2008,*belesi2010,*miller2010} However, unlike conventional ferrimagnets and unlike Cu$_2$OSeO$_3$, BaAg$_2$Cu[VO$_4]_2$ does not feature well-defined sublattices with opposite directions of the spin, and rather shows a sequence of FM and AFM layers. Unfortunately, the frustrated nature of the interlayer coupling $J_c$ prevents us from using QMC for simulating the ground-state properties and the transition temperature $T_C$. Therefore, we are presently unable to verify the proposed magnetic structure. Further experimental studies, such as neutron diffraction, would be required to tackle this problem.

The microscopic magnetic model of BaAg$_2$Cu[VO$_4]_2$ is furthermore instructive from a structural viewpoint. The Cu1 and Cu2 sites look deceptively similar, so that one would not expect any substantial difference between the magnetic couplings within the two sublattices. However, the couplings are very different -- not only in the magnitude but also in the nature -- because of the subtle influence of the VO$_4$ tetrahedra connecting the neighboring CuO$_4$ plaquettes. The effect of the non-magnetic group is sizable and two-fold. Vanadium $3d$ orbitals contribute to the Wannier functions, and induce a FM superexchange, which is weakly dependent on the specific arrangement of the VO$_4$ tetrahedra. This FM contribution represents a constant term that is superimposed on a variable AFM superexchange. The latter is controlled by the Cu--Cu hoppings showing dramatic dependence on the mutual orientation of the VO$_4$ tetrahedra and CuO$_4$ plaquettes. Depending on the specific geometry, the AFM contributions may or may not surpass the FM superexchange, and qualitatively different exchange couplings emerge.

The subtle dependence of AFM superexchange on the orientation of non-magnetic tetrahedra is reminiscent of the long-range couplings in BiCu$_2$PO$_6$, where slight rotations of the bridging PO$_4$ groups modify the interactions by $50-70$~K.\cite{tsirlin2010} More generally, the unusual microscopic scenario of BaAg$_2$Cu[VO$_4]_2$ confirms the crucial importance of non-magnetic bridging groups for the superexchange in magnetic insulators. Other remarkable examples include the effect of GeO$_4$ tetrahedra on the Cu-based spin chains in CuGeO$_3$ (Ref.~\onlinecite{geertsma1996}), as well as the unusual ferromagnetism of CdVO$_3$ related to the low-lying $5s$ orbitals of Cd atoms.\cite{cdvo3} The effect of the non-magnetic groups opens broad prospects for tweaking superexchange couplings by minor alterations of the crystal structure. For example, BaAg$_2$Cu[VO$_4]_2$ is likely to sustain cation substitutions in the Ba and Ag positions, thus leading to further interesting combinations of FM and/or AFM spin chains in a single chemical compound.

In summary, we have derived a microscopic magnetic model of BaAg$_2$Cu[VO$_4]_2$, and presented a consistent interpretation of the available experimental data for this compound. The crucial and highly unexpected feature of BaAg$_2$Cu[VO$_4]_2$ is the dramatic difference between the couplings within the Cu1 and Cu2 sublattices. While the Cu1 sublattice is ferromagnetic, the Cu2 sublattice is antiferromagnetic. This unusual -- and so far unreported -- combination of weakly coupled FM and AFM spin chains within a single chemical compound leads to peculiar thermodynamic properties, with the specific heat resembling that of a strongly frustrated two-dimensional spin system. The spin lattice of BaAg$_2$Cu[VO$_4]_2$ is, however, only weakly frustrated and quasi-1D, as confirmed by the high-field magnetization measurements suggesting the ground state with non-zero net magnetization. The different couplings within similar structural units are solely determined by the orientation of the non-magnetic VO$_4$ tetrahedra with respect to the CuO$_4$ plaquettes. These results present an instructive example on the importance of bridging groups for superexchange pathways, and open interesting opportunities for tuning low-dimensional spin systems within a given structure type.

\acknowledgments
We are grateful to Oleg Janson and Deepa Kasinathan for stimulating discussions, and to Ngozi E. Amuneke for her help in sample preparation. The high-field magnetization measurements were supported by EuroMagNET~II under the EC contract 228043. A.T. acknowledges the funding from Alexander von Humboldt Foundation. A.M. appreciates the support from the Welch Foundation (Grant G099857).

%

\end{document}